\documentclass[preprint2]{aastex}

\shorttitle{Discovery of VHE Gamma-rays from W\,Comae}
\shortauthors{M.Beilicke et al. (VERITAS collaboration)}

\begin{document}


\title{VERITAS Discovery of $\mathbf >$200~GeV Gamma-ray Emission from
the Intermediate-frequency-peaked BL\,Lac Object W\,Comae}


\author{
V.A. Acciari\altaffilmark{20,1},
E. Aliu\altaffilmark{23},
M. Beilicke\altaffilmark{2,*},
W. Benbow\altaffilmark{1},
%%%%%[-- G. Blaylock\altaffilmark{3},
M. B\"ottcher\altaffilmark{27},
S.M. Bradbury\altaffilmark{4},
J.H. Buckley\altaffilmark{2},
V. Bugaev\altaffilmark{2},
Y. Butt\altaffilmark{24},
%%%%%[-- K.L. Byrum\altaffilmark{5},
O. Celik\altaffilmark{6}, 
A. Cesarini\altaffilmark{1,21},
L. Ciupik\altaffilmark{7},
Y.C.K. Chow\altaffilmark{6},
P. Cogan\altaffilmark{12},
P. Colin\altaffilmark{11},
W. Cui\altaffilmark{8},
M.K. Daniel\altaffilmark{4,\dagger},
%%%%%[-- P. Dowkontt\altaffilmark{2},
%%%%%[-- C. Duke\altaffilmark{16},
T. Ergin\altaffilmark{3},
A.D. Falcone\altaffilmark{22},
%%%%%[-- D.J. Fegan\altaffilmark{13},
S.J. Fegan\altaffilmark{6},
J.P. Finley\altaffilmark{8},
G. Finnegan\altaffilmark{11},
P. Fortin\altaffilmark{14},
L.F. Fortson\altaffilmark{7},
A. Furniss\altaffilmark{17},
D. Gall\altaffilmark{8},
%%%%%[-- K. Gibbs\altaffilmark{1},
G.H. Gillanders\altaffilmark{21},
J. Grube\altaffilmark{13},
R. Guenette\altaffilmark{12},
G. Gyuk\altaffilmark{7},
%%%%%[-- J. Hall\altaffilmark{25},
D. Hanna\altaffilmark{12},
E. Hays\altaffilmark{5,1},
J. Holder\altaffilmark{23},
D. Horan\altaffilmark{5},
%%%%%[-- S.B. Hughes\altaffilmark{2},
C.M. Hui\altaffilmark{11},
T.B. Humensky\altaffilmark{10},
A. Imran\altaffilmark{9},
P. Kaaret\altaffilmark{18},
N. Karlsson\altaffilmark{7},
%%%%%[-- G.E. Kenny\altaffilmark{12},
M. Kertzman\altaffilmark{15},
D.B. Kieda\altaffilmark{11},
%%%%%[-- J. Kildea\altaffilmark{1},
A. Konopelko\altaffilmark{8},
H. Krawczynski\altaffilmark{2},
F. Krennrich\altaffilmark{9},
M.J. Lang\altaffilmark{21},
S. LeBohec\altaffilmark{11},
K. Lee\altaffilmark{2},
G. Maier\altaffilmark{12},
%%%%%[-- H. Manseri\altaffilmark{11},
A. McCann\altaffilmark{12},
M. McCutcheon\altaffilmark{12},
%%%%%[-- J. Millis\altaffilmark{8},
P. Moriarty\altaffilmark{20},
R. Mukherjee\altaffilmark{14},
T. Nagai\altaffilmark{9},
J. Niemiec\altaffilmark{9,\ddagger},
%%%%%[-- P.A. Ogden\altaffilmark{4},
R.A. Ong\altaffilmark{6},
D. Pandel\altaffilmark{18},
J.S. Perkins\altaffilmark{1},
D. Petry\altaffilmark{26},
%%%%%[-- F. Pizlo\altaffilmark{8},
M. Pohl\altaffilmark{9},
J. Quinn\altaffilmark{13},
K. Ragan\altaffilmark{12},
L.C. Reyes\altaffilmark{10},
P.T. Reynolds\altaffilmark{19},
E. Roache\altaffilmark{1},
H.J. Rose\altaffilmark{4},
M. Schroedter\altaffilmark{9},
G.H. Sembroski\altaffilmark{8},
A.W. Smith\altaffilmark{1,4},
D. Steele \altaffilmark{7},
S.P. Swordy\altaffilmark{10},
%%%%%[-- A. Syson\altaffilmark{4},
J.A. Toner\altaffilmark{1,21},
%%%%%[-- L. Valcarcel\altaffilmark{12},
V.V. Vassiliev\altaffilmark{6},
R. Wagner\altaffilmark{5},
S.P. Wakely\altaffilmark{10},
J.E. Ward\altaffilmark{13},
T.C. Weekes\altaffilmark{1},
A. Weinstein\altaffilmark{6},
R.J. White\altaffilmark{4},
D.A. Williams\altaffilmark{17},
S.A. Wissel\altaffilmark{10},
M. Wood\altaffilmark{6},
B. Zitzer\altaffilmark{8}
}


\altaffiltext{1}{Fred Lawrence Whipple Observatory, Harvard-Smithsonian 
Center for Astrophysics, Amado, AZ 85645, USA}

\altaffiltext{2}{Department of Physics, Washington University, St. 
Louis, MO 63130, USA}

\altaffiltext{3}{Department of Physics, University of Massachusetts, 
Amherst, MA 01003-4525, USA}

\altaffiltext{4}{School of Physics and Astronomy, University of Leeds, 
Leeds LS2 9JT, UK}

\altaffiltext{5}{Argonne National Laboratory, 9700 S. Cass Avenue, 
Argonne, IL 60439, USA}

\altaffiltext{6}{Department of Physics and Astronomy, University of 
California, Los Angeles, CA 90095, USA}

\altaffiltext{7}{Astronomy Department, Adler Planetarium and Astronomy 
Museum, Chicago, IL 60605, USA}

\altaffiltext{8}{Department of Physics, Purdue University, West 
Lafayette, IN 47907, USA}

\altaffiltext{9}{Department of Physics and Astronomy, Iowa State 
University, Ames, IA 50011, USA}

\altaffiltext{10}{Enrico Fermi Institute, University of Chicago, 
Chicago, IL 60637, USA}

\altaffiltext{11}{Physics Department, University of Utah, Salt Lake 
City, UT 84112, USA}

\altaffiltext{12}{Physics Department, McGill University, Montreal, QC 
H3A 2T8, Canada}

\altaffiltext{13}{School of Physics, University College Dublin, 
Belfield, Dublin, Ireland }

\altaffiltext{14}{Department of Physics and Astronomy, Barnard College, 
Columbia University, NY 10027, USA}

%\altaffiltext{15}{Department of Physics and Astronomy, DePauw University, Greencastle, IN 46135-0037, USA}

\altaffiltext{16}{Department of Physics, Grinnell College, Grinnell, IA 
50112-1690, USA}

\altaffiltext{17}{Santa Cruz Institute for Particle Physics and 
Department of Physics, University of California, Santa Cruz, CA 95064, 
USA}

\altaffiltext{18}{Department of Physics and Astronomy, University of 
Iowa, Van Allen Hall, Iowa City, IA 52242, USA}

\altaffiltext{19}{Department of Applied Physics and Instrumentation, 
Cork Institute of Technology, Bishopstown, Cork, Ireland}

\altaffiltext{20}{Department of Life and Physical Sciences, Galway-Mayo 
Institute of Technology, Dublin Road, Galway, Ireland}

\altaffiltext{21}{Physics Department, National University of Ireland, 
Galway, Ireland}

\altaffiltext{22}{Department of Astronomy and Astrophysics, Penn State 
University, University Park, PA 16802, USA}

\altaffiltext{23}{Department of Physics and Astronomy, Bartol Research 
Institute, University of Delaware, Newark, DE 19716, USA}

\altaffiltext{24}{Smithsonian Astrophysical Observatory, Cambridge, MA 
02138, USA}

%\altaffiltext{25}{Fermi National Accelerator Laboratory, Batavia, IL 60510, USA}

\altaffiltext{26}{Max-Planck Institut for Extraterrestrial Physics 
(MPE), Giessenbachstrasse, 85748 Garching, Germany}

\altaffiltext{27}{Department of Physics and Astronomy Astrophysical
Institute, Clippinger 339, Ohio University, Athens, OH 45701 - 2979}

\altaffiltext{$\dagger$}{Now at: Department of Physics, Durham 
University, South Road, Durham, DH1 3LE, U.K.} 

\altaffiltext{$\ddagger$}{Now at: Instytut Fizyki J\c{a}drowej PAN, ul.  
Radzikowskiego 152, 31-342 Krak\'ow, Poland} 

\altaffiltext{*}{Corresponding author: beilicke@physics.wustl.edu}


\begin{abstract}

We report the detection of very high-energy $\gamma$-ray emission from
the intermediate-frequency-peaked BL\,Lacertae object W\,Comae ($z =
0.102$) by VERITAS, an array of four imaging atmospheric-Cherenkov
telescopes. The source was observed between January and April 2008. A
strong outburst of $\gamma$-ray emission was measured in the middle of
March, lasting for only four days. The energy spectrum measured during
the two highest flare nights is fit by a power-law and is found to be
very steep, with a differential photon spectral index of $\Gamma = 3.81
\pm 0.35_{\rm{stat}} \pm 0.34_{\rm{syst}}$. The integral photon flux
above $200 \, \rm{GeV}$ during those two nights corresponds to roughly
$9\%$ of the flux from the Crab Nebula. Quasi-simultaneous Swift
observations at X-ray energies were triggered by the VERITAS
observations. The spectral energy distribution of the flare data can be
described by synchrotron-self-Compton (SSC) or external-Compton (EC)
leptonic jet models, with the latter offering a more natural set of
parameters to fit the data.

\end{abstract}



\section{Introduction}

The blazars detected at very high energies (VHE, $E > 100 \, \rm{GeV}$)
by ground-based imaging atmospheric-Cherenkov telescopes (IACTs) are
extreme objects in the active galactic nuclei (AGN) population. 
Typically these sources show core-dominated emission, and they are
characterized by rapid variability and strong broadband continuum
emission ranging from the radio band to the X-ray band. 
Multi-wavelength data on blazars reveal that their spectral energy
distribution (SED) is characterized by two broad, well-separated
``humps'' arising from synchrotron (low-energy) and inverse-Compton (IC)
or hadronic emission (high-energy). Blazars are categorized into
different sub-classes based on the frequencies at which these emission
components reach a maximum. Flat-spectrum radio quasars (FSRQs) and
low-frequency-peaked BL\,Lacs (LBLs) are generally seen to have
low-frequency, synchrotron peaks in the IR/optical regime, whereas
high-frequency-peaked BL\,Lacs (HBLs) exhibit peaks in the X-ray band,
in several cases at energies of $\sim$100~keV or higher.
Intermediate-frequency-peaked BL\,Lacs (IBLs) bridge the gap between
LBLs and HBLs. The properties of the broad sub-classes of blazars, the
luminosity-versus-frequency trends and possible physical explanations
are discussed by \cite{ghi08}.

Gamma rays are an important component of the SED of blazars; the
integral power in the $\gamma$-ray waveband is comparable or higher than
that in the rest of the electromagnetic spectrum (from radio to X-rays).
There are $65$ blazars detected at MeV/GeV energies by the EGRET
instrument on board the {\sl Compton Gamma Ray Observatory (CGRO)}
\citep{mat01, har99} and several of the other EGRET sources also have
likely blazar counterparts. Ground-based IACTs have established $\sim$20
blazars as emitters of VHE $\gamma$-radiation; see for example
\citet{wak07}. While the blazars detected at MeV/GeV energies tend to be
largely FSRQs (and some LBLs), almost all VHE blazars belong to the
class of HBLs, the only exceptions being the LBLs BL\,Lacertae
\citep{alb07}, S5\,0716+71 \citep{tes08} and the FSRQ 3C\,279
\citep{tes07}.

The IBL W\,Comae (W\,Com) at a redshift of $z = 0.102$ has long been an
object of interest for VHE observatories. W\,Com was discovered at radio
frequencies \citep{bir71} and later detected at X-ray energies by the
{\sl Einstein} Imaging Proportional Counter in June 1980 \citep{wor90}.
Data taken with the {\sl BeppoSAX} satellite in 1998 \citep{tag00}
clearly showed that the transition between the low- and high-energy
peaks in the SED occurs around $\sim$4~keV. In April-May 1998, an
exceptional optical outburst was detected from W\,Com showing rapid
variability on time scales of hours \citep{mas99}.

At $\gamma$-ray energies, W\,Com was detected by EGRET in the $100 \,
\rm{MeV} - 10 \, \rm{GeV}$ band \citep{har99} and in a re-analysis of
the data up to $25 \, \rm{GeV}$ \citep{din01}. Due to its rather hard
EGRET spectrum (photon spectral index $\alpha=1.73\pm 0.18$), with no
sign of spectral cut-off \citep{har99}, the source became even more
interesting for VHE observations. However, W\,Com was not detected by
the Whipple IACT above $300 \, \rm{GeV}$ in 1993/94 \citep{ker95} and
1995/96/98 \citep{hor04}, nor by the solar heliostat Cherenkov telescope
STACEE \citep{sca04}. In this paper we report the discovery of VHE
$\gamma$-ray emission from W\,Com with VERITAS.

%%%%%%%%%%%%%%%%%%%%%%%%%%%%%%%%%%%%%%%%%%%%%%%%%%%%%%%%%%%%%%%%%%%%%%%%%
%%%%%%%%%%%%%%%%%%%%%%%%%%%%%%%%%%%%%%%%%%%%%%%%%%%%%%%%%%%%%%%%%%%%%%%%%
%%%%%%%%%%%%%%%%%%%%%%%%%%%%%%%%%%%%%%%%%%%%%%%%%%%%%%%%%%%%%%%%%%%%%%%%%
\section{VERITAS and Swift Observations and Results}

%%%%%%%%%%%%%%%%%%%%%%%%%%%%%%%%%%%%%%%%%%%%%%%%%%%%%%%%%%%%%%%%%%%%%%%

\begin{figure}[t]
\epsscale{0.99}
\plotone{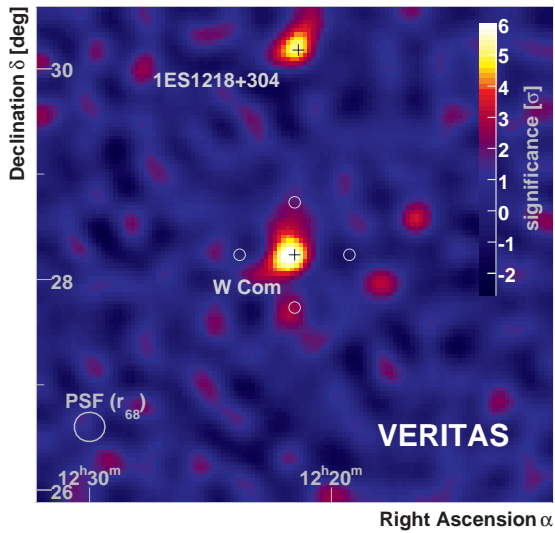}

\caption{\label{fig:SkyMap} Sky map of significances of the whole data
set ({\it standard cuts}, oversampling radius of $0.15\degr$). The
background is estimated using the ring background model
\citep{ber07}. The excess $\sim$2$\degr$ North of W\,Com (cross)
corresponds to 1ES\,1218+304 (cross) and demonstrates the capability of
VERITAS to detect sources at the edge of the field of view of the
camera. The telescope tracking positions in wobble mode (open circles)
and the angular resolution (PSF) are indicated as well.}

\end{figure}

VERITAS consists of four $12 \, \rm{m}$ diameter IACTs and is located at
the basecamp of the Fred Lawrence Whipple Observatory (FLWO) in southern
Arizona at an altitude of $1280 \, \rm{m}$. It detects the Cherenkov
light emitted by an extensive air shower (initiated by a VHE
$\gamma$-ray photon or cosmic ray entering the Earth's atmosphere) using
a 499-pixel photomultiplier camera located in the focal plane of each
telescope. The array is sensitive to $\gamma$-rays in the energy range
from $\sim$100~GeV to $\sim$30~TeV. Observations are performed in
moonless nights in the ``wobble'' mode of operation, where the
telescopes are pointed to positions offset by $\pm 0.5 \degr$
(alternating in direction) with respect to the source position, to allow
for a simultaneous background estimation. More details about VERITAS,
the data calibration and the analysis techniques can be found in
\citet{acc08}.

Only shower images which pass certain quality cuts are considered in the
event reconstruction (image size $\geq 500$ digital counts
(dc)\footnote{The photomultiplier pulses are integrated within a time
window of $24 \, \rm{ns}$ duration. One digital count corresponds to
approximately $5$ photoelectrons.}; distance between the image center of
gravity and the center of the camera $\leq 1.43 \degr$). The
$\gamma$/hadron separation cuts used in this analysis are based on the
width and length of the recorded images \citep{acc08} and were optimized
{\it a priori} on Crab Nebula data for a source with a flux at the
$5\%$-level of the Crab Nebula. We refer to these as {\it standard
cuts}. An event is considered to fall into the signal (ON) region, if
the squared angular distance $\Delta \theta^{2}$ between the
reconstructed event direction and the W\,Com position is less than
$0.0125 \, \rm{deg}^{2}$. The background is estimated from different
regions of equal size positioned at the same radial distance from the
camera center as the ON region \citep{ber07}. This background model,
referred to as the ``reflected region model'', is used unless otherwise
stated. Since the energy spectrum of W\,Com is found to be very steep
(see below) a second set of cuts (optimized on Crab Nebula data for low
energies of $E \leq 200 \, \rm{GeV}$) is used to derive the energy
spectrum and the light curve. These {\it a posteriori} cuts are referred
to as {\it soft cuts} in this paper and use an image size $\geq 250 \,
\rm{dc}$ and an angular distance to the source position of $\Delta
\theta^{2} \leq 0.02 \, \rm{deg}^{2}$. All results obtained with the
{\it soft cuts} are in good agreement with the ones obtained using the
{\it standard cuts}.

VERITAS observed W\,Com from January to April 2008 for a total of
$39.5$~hours (deadtime corrected) after run quality selection. The
zenith-angle range of the observations was $3\degr - 45\degr$ with an
average of $19\degr$, corresponding to an analysis energy
threshold\footnote{The energy threshold is defined as the energy
corresponding to the peak detection rate for a Crab-like spectrum.} of
$260 \, \rm{GeV}$ ({\it standard cuts}) and $180 \, \rm{GeV}$ ({\it soft
cuts}). In the entire data set, $111$ excess events ($543$ ON events and
$432$ normalized OFF events, normalization $\alpha = 0.111$) are
detected from the direction of W\,Com using the {\it standard} cuts,
corresponding to a statistical significance of $4.9$~standard deviations
($4.9 \, \sigma$), calculated following \citet{li83}. The sky map
showing W\,Com and the known VHE blazar 1ES\,1218+304\footnote{A
significant VHE $\gamma$-ray excess was recorded from 1ES\,1218+304
during the dedicated W\,Com observations which will be addressed in a
forthcoming paper.} \citep{alb06, for07} located in the same field of
view of the dedicated W\,Com observations is shown in
Fig.~\ref{fig:SkyMap}. The mean position of the W\,Com excess is derived
by fitting a 2D Gaussian function to the uncorrelated excess sky map and
is found to be compatible within errors with the nominal position of
W\,Com. 

\begin{figure}[t]
\epsscale{0.8}
\plotone{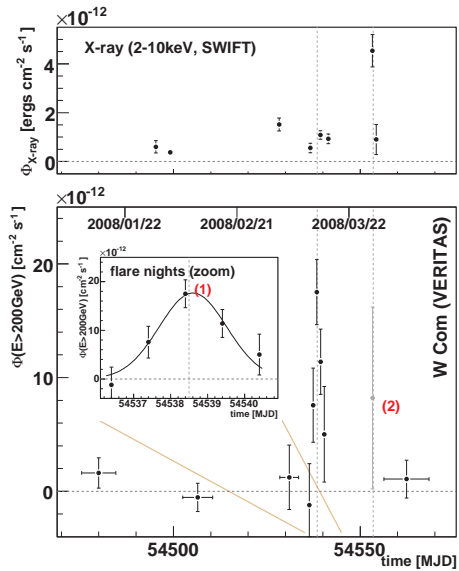}

\caption{\label{fig:LC} {\bf Lower panel:} The light curve $I(E>200 \,
\rm{GeV})$ is shown ({\it soft} cuts, assuming a spectral shape of
$\rm{d}N/\rm{d}E \propto E^{-\Gamma}$ with $\Gamma = 3.8$). Each flux
point corresponds to one observation period (defined by $\sim$3 weeks of
operation between two full-moon phases), with the exception of the flare
around MJD~54538 (1) for which a night-by-night binning is used (see
inlay for details; the fitted model light curve is described in the
text). {\bf Upper panel:} The X-ray flux as measured by Swift for the
same time period. The vertical lines are shown for easier comparison. 
The simultaneous VERITAS/Swift measurements around MJD 54553.3 (2)
during a high X-ray flux level is discussed in the text.} 

\end{figure}

Almost the entire excess from W\,Com ($>70\%$) is recorded during a
strong flare, which occured during four nights in the middle of March
\citep{swo08}; see Fig.~\ref{fig:LC}. The measured excess of the whole
corresponding observation period~-- modified Julian date (MJD) 54528.4
to 54540.4~-- corresponds to a statistical significance of $6.3 \,
\sigma$ ({\it standard cuts}, $85$ excess events) and $8.6 \, \sigma$
({\it soft cuts}, $275$ excess events). Correcting for $8$ trials (four
observations periods and two sets of cuts) results in $5.9 \, \sigma$
and $8.3 \, \sigma$, respectively. No statistically significant excess
is measured in the remaining data set. A fit of a constant function to
the whole night-by-night light curve (January to April) results in a
probability of constant emission of $2.1 \cdot 10^{-4}$. In order to
estimate the time scale of the flux variations the light curve of the
flare nights (see inlay of Fig.~\ref{fig:LC}) is modeled by the function
$\Phi(t) = \Phi_{0} \times \exp \left( -(t-t_{0})^{2} /
\sigma_{\rm{t}}^{2} \right)$ with the flare occurring at $t_{0} =
54538.6 \pm 0.2 \, \rm{MJD}$ with the characteristic time scale of
$\sigma_{\rm{t}} = 1.29 \pm 0.28 \, \rm{days}$. No significant flux
variations are measured within individual nights.

\begin{figure}[t]
\epsscale{0.99}
\plotone{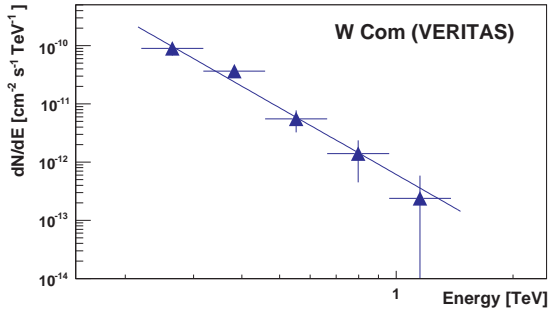}

\caption{\label{fig:Spectrum} Differential energy spectrum of W\,Com
({\it soft cuts}), derived from the two highest flare nights: label (1)
in Fig.~\ref{fig:LC}. The parameters of the fitted power-law function
(line) are summarized in the text.}

\end{figure}

A differential energy spectrum is derived for the two highest flare
nights. The spectrum is shown in Fig.~\ref{fig:Spectrum} and is well fit
($\chi^{2}/\rm{dof} = 2.9/3$) by a power-law function $\rm{d}N/\rm{d}E =
I_{0} \times (E/400\,\rm{GeV})^{-\Gamma}$, with $I_{0} = (2.00 \pm
0.31_{\rm{st}}) \times 10^{-11} \, \rm{cm}^{-2} \, \rm{s}^{-1} \,
\rm{TeV}^{-1}$ and $\Gamma = 3.81 \pm 0.35_{\rm{st}}$. The integral
photon flux above $200 \, \rm{GeV}$ is calculated to be
$\Phi_{E>200\,\rm{GeV}} = (1.99 \pm 0.07_{\rm{st}}) \times 10^{-11} \,
\rm{cm}^{-2} \, \rm{s}^{-1}$, corresponding to $9\%$ of the flux
measured from the Crab Nebula above the same energy. The systematic
errors on the normalization constant and the photon index for this
low-energy regime are estimated to be $\Delta I_{0} / I_{0} = 25\%$ and
$\Delta \Gamma / \Gamma = 9\%$, respectively.

Simultaneous Swift observations of W\,Com were performed for a total
duration of $11.6 \, \rm{h}$. Swift comprises a UV instrument UVOT and
X-ray instruments XRT and BAT \citep{geh04}. Data reduction and
calibration are performed with the {\it HEAsoft 6.4}
package\footnote{\url{http://heasarc.gsfc.nasa.gov/lheasoft/}} and the
{\it XRTPIPELINE} tool. All XRT data presented here were taken in Photon
Counting (PC) mode. Standard filtering criteria are applied. Photon
pile-up effects are negligible in the data. All energy spectra are fit
with an absorbed power law using {\it XSPEC 12.4}. A galactic column
density of $N_{\rm{H}} = 1.88 \cdot 10^{20} \, \rm{cm}^{2}$ was assumed
\citep{dic90}. No significant deviation from a power law spectral shape
is found within the limited statistics. UVOT observations were taken
over the six photometric bands of V, B, U, UVW1, UVM2, and UVW2
\citep{poo08}. The UVOTSOURCE tool is used to extract counts, correct
for coincidence losses, apply background subtraction and calculate the
source flux. The source fluxes are de-reddened using the interstellar
extinction curve in \citet{fit99}.

The light curve of the X-ray flux is shown in Fig.~\ref{fig:LC}, upper
panel. No change in spectral slope could be detected when comparing
results for individual nights. An X-ray flux at a level roughly $4$
times higher than the flux observed during the VHE flare was observed
around MJD 54553.3 (see Fig.~\ref{fig:LC}, top panel). VERITAS also
observed W\,Com during this night for $\sim$40~min but the data do not
pass the standard quality selection\footnote{Showing a cosmic ray
triggerrate $\sim$27$\%$ lower than expected due to non-optimal weather
conditions~-- with the maximum allowed deviation being $20\%$.}. 
Nevertheless, since the VERITAS data (MJD 54553.3) are simultanenous
with the X-ray flare the flux derived from these data (including an
additional $50\%$ systematical error) is shown for reference in
Fig.~\ref{fig:LC}, label (2). The $99.9 \% \, \rm{c.l.}$ upper limit
(assuming an underestimation of the count rates by $50\%$) is calculated
to be $\sim$2~times higher than the peak flux measured during the VHE
flare. Although no detailed conclusions can be drawn, a linear X-ray/TeV
flux correlation does not seem likely.

%%%%%%%%%%%%%%%%%%%%%%%%%%%%%%%%%%%%%%%%%%%%%%%%%%%%%%%%%%%%%%%%%%%%%%%

%%%%%%%%%%%%%%%%%%%%%%%%%%%%%%%%%%%%%%%%%%%%%%%%%%%%%%%%%%%%%%%%%%%%%%%%%
%%%%%%%%%%%%%%%%%%%%%%%%%%%%%%%%%%%%%%%%%%%%%%%%%%%%%%%%%%%%%%%%%%%%%%%%%
%%%%%%%%%%%%%%%%%%%%%%%%%%%%%%%%%%%%%%%%%%%%%%%%%%%%%%%%%%%%%%%%%%%%%%%%%
\section{Modeling and Discussion}

\begin{figure}[t]
\epsscale{0.99}
\plotone{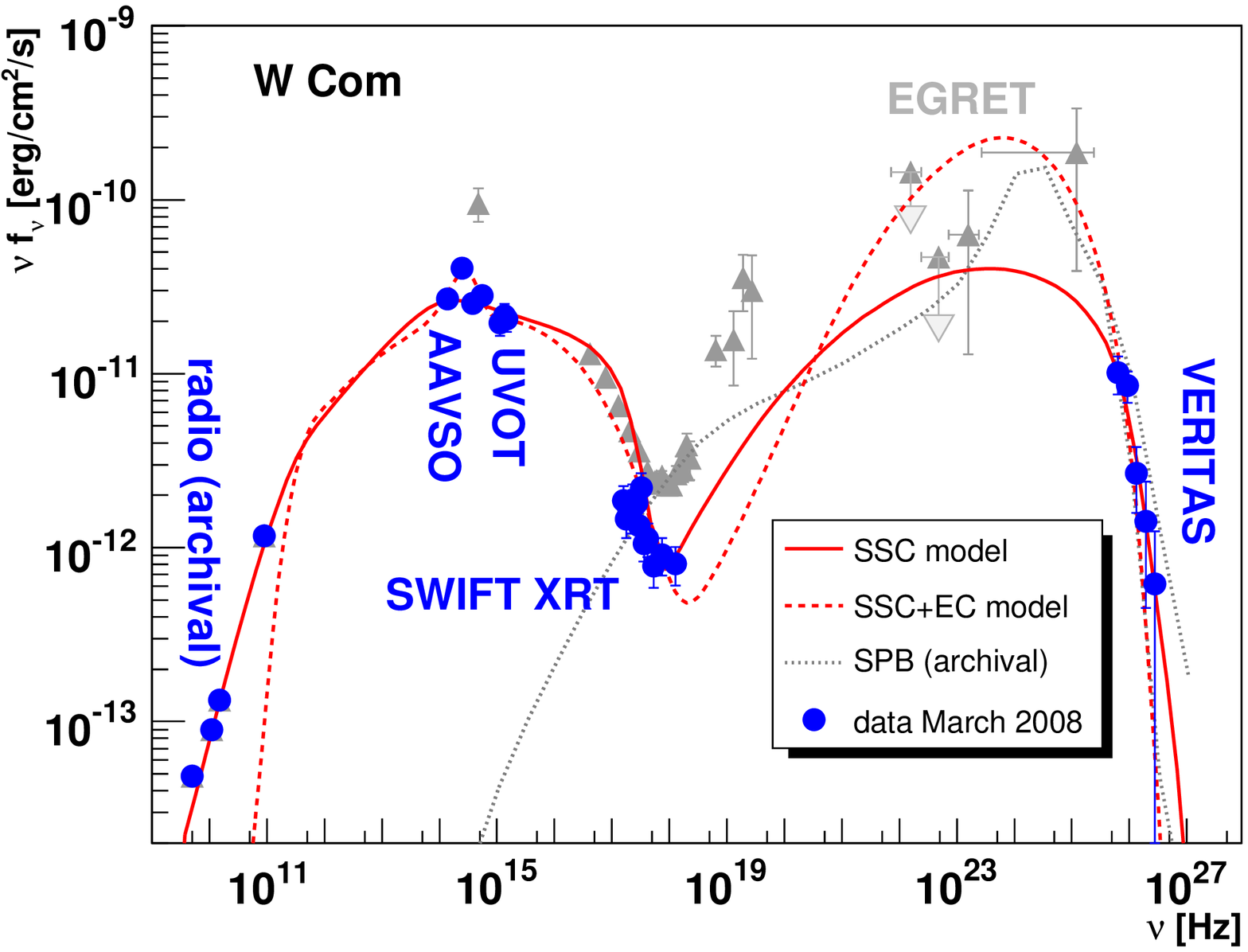}

\caption{\label{fig:SED} Quasi-simultaneous SED of W\,Com, including the
VERITAS flare data (MJD 54538.4 and 54539.4, see
Fig.~\ref{fig:Spectrum}), the Swift XRT/UVOT data (MJD 54539.4) and the
V and I band data (MJD 54540) from AAVSO \citep{bed08}.  The radio data
are non-simultaneous and the same as in \citet{boe02a}. Details of the
SSC and SSC+EC model fits (see legend) are described in the text. The
hadronic SPB model curve~1 from \citet{boe02a} is shown for reference.
Archival data (optical, X-ray and 1998 EGRET data) are shown as grey
points for comparison, see references in \citet{boe02a}.}

\end{figure}

The VERITAS data taken at MJD 54538.4 and 54539.4 are used to model the
SED of W\,Com (see Fig.~\ref{fig:SED}) together with the simultaneous
SWIFT XRT/UVOT (MJD 54539.4) and optical AAVSO data (MJD 54540,
\citet{bed08}), as well as archival radio data. The following model
curves are corrected for $\gamma \gamma$ absorption by the extragalactic
background light according to the ``best fit'' model of \citet{kne04}.
The SED can be fit by a simple one-zone SSC model, using the equilibrium
version of the code of \citet{boe02b}. Here an ad-hoc non-thermal
electron injection spectrum with particle index $q$ and total particle
injection luminosity $L_{\rm{inj}}$ is balanced self-consistently with
radiative cooling from synchrotron and Compton emission. The best fit to
the SED is shown in Fig.~\ref{fig:SED} as a solid line. The parameters
of the fit are: $\gamma_{1} = 450$, $\gamma_{2} = 4.5 \cdot 10^{5}$, $q
= 2.2$, $L_{\rm{inj}} = 2.8 \cdot 10^{45} \, \rm{erg}/\rm{s}$, a
magnetic field of $B = 0.007 \, \rm{G}$, a doppler factor of $\delta =
30$ and a size of the emission region of $R = 10^{17} \, \rm{cm}$. The
wide separation of the SED peaks, together with the very low X-ray flux,
require an unusually low magnetic field in order to allow for
sufficiently high particle Lorentz factors to produce the observed VHE
$\gamma$-ray flux. The ratio between the magnetic and electron energy
densitiy is $\sigma = 1.3 \cdot 10^{-3}$. The light crossing time $\tau
= R/(c \cdot \delta) \approx 1.3 \, \rm{d}$ matches the time scale
observed in the VHE flare (compare Fig.~\ref{fig:LC}), but it is
relatively large compared with the extremely rapid VHE variability on
time scales of $2$ to $10$~minutes seen in other TeV blazars at higher
flux levels \citep{alb07b, aha07}. 

The SED was also fit by a self-consistent model that contains both SSC
emission and an EC component, similar to the model of \citep{ino96}. The
external photons are assumed as steady-state blackbody radiation peaking
in the near-infrared (radius 1800 Schwarzschild radii, $0.4\%$ of the
Eddington Luminosity). The particles are accelerated by diffusive shock
acceleration and the maximum electron Lorentz factor $\gamma_{\rm{max}}$
is determined by competition between acceleration and radiative cooling.
As for the SSC fit, a cooling break in the electron spectrum is assumed
to occur at the energy where the cooling time becomes shorter than the
light crossing time of the emission region. Finally, we assume that the
electron distribution has some minimum Lorentz factor $\gamma_{\rm min}$
from some unknown injection process. The power-law slope of the electron
spectrum (without the cooling break) is parameterized by
$\rm{d}N/\rm{d}E \propto E^{-s}$, where the free parameter $s$ is
expected to vary between $2.3$ (for canonical first-order Fermi
ultra-relativistic shock acceleration) and $2.0$ (for canonical
non-relativistic first order-Fermi acceleration by a strong shock).
However, it should be noted that relativistic shocks could produce much
harder energy spectra \citep{ste07}.

A reasonably good fit (see Fig.~\ref{fig:SED}) is obtained taking $B=0.3
\, \rm{G}$, $\delta = \Gamma = 30$, $\sigma=1.0$ (assuming
equipartition), $\gamma_{\rm min} = \Gamma = 30$ and $R=1.76 \times
10^{16} \, {\rm cm}$. To match the inferred shape of the electron
spectrum, rather inefficient particle acceleration is invoked (e.g.,
with $u_{\rm shock} = 0.1\, c$ in the bulk frame and the ratio of the
electron scattering mean-free-path to the Bohm limit of $3000$) with an
electron spectrum with index $s = 2.0$. For this choice of parameters,
the model gives an acceleration time (equal to the cooling time) at the
maximum electron energy of $7.2\, {\rm min}$. Assuming that the emission
region of radius $R$ is comoving with the jet, the light crossing time
for these parameters is $\tau = 330\, {\rm min}$. This value is closer
to the typical variability time scales of other VHE blazars and
consistent with our observed lightcurve.

The synchrotron proton blazar (SPB) model from \citet{boe02a} fitted to
data of the 1998 W\,Com campaign is also shown for reference in
Fig.~\ref{fig:SED}.

%%%%%%%%%%%%%%%%%%%%%%%%%%%%%%%%%%%%%%%%%%%%%%%%%%%%%%%%%%%%%%%%%%%%%%%

%%%%%%%%%%%%%%%%%%%%%%%%%%%%%%%%%%%%%%%%%%%%%%%%%%%%%%%%%%%%%%%%%%%%%%%%%
%%%%%%%%%%%%%%%%%%%%%%%%%%%%%%%%%%%%%%%%%%%%%%%%%%%%%%%%%%%%%%%%%%%%%%%%%
%%%%%%%%%%%%%%%%%%%%%%%%%%%%%%%%%%%%%%%%%%%%%%%%%%%%%%%%%%%%%%%%%%%%%%%%%
\section{Summary \& Conclusion}

VERITAS detected VHE $\gamma$-ray emission from W\,Com with a
statistical significance of $4.9$~standard deviations for the entire
data set (January to April 2008). A strong outburst was observed in
March 2008 with a statistical significance of $>8$~standard deviations,
that lasted for only four days. In addition to W\,Com, a second
extragalctic source (the VHE blazar 1ES\,1218+304) is detected in the
same field of view -- for the first time in VHE $\gamma$-ray astronomy.

W\,Com is the first VHE-detected blazar of the IBL class. The extension
of the VHE catalog to the FRSQ, LBL and IBL classes will play a major
role in our understanding of blazar populations and dynamics. The
quasi-simultaneous SED of W\,Com at the time of the VHE outburst can be
modeled with a simple one-zone SSC model. However, an unusually low
magnetic field of $B = 0.007 \, \rm{G}$ (more than an order of magnitude
lower than typically found in the modeling of other BL\,Lac-type
blazars) and a small ratio of the magnetic field to electron energy
density of $\sigma = 1.3 \cdot 10^{-3}$ are required. An EC model with
more natural parameters ($B=0.36 \, \rm{G}$ and $\sigma = 1$) provides a
good fit and could account for shorter variability time scales. Our
model results agree with the expectation that for IBLs (and LBLs) the
higher optical luminosity plays an important role in providing the seed
population for IC scattering.

The IBL W\,Com will be an excellent target for future observations at
GeV energies with GLAST and in the VHE regime with IACTs, including
correlated GeV/TeV variability studies.

%%%%%%%%%%%%%%%%%%%%%%%%%%%%%%%%%%%%%%%%%%%%%%%%%%%%%%%%%%%%%%%%%%%%%%%%%
%%%%%%%%%%%%%%%%%%%%%%%%%%%%%%%%%%%%%%%%%%%%%%%%%%%%%%%%%%%%%%%%%%%%%%%%%
%%%%%%%%%%%%%%%%%%%%%%%%%%%%%%%%%%%%%%%%%%%%%%%%%%%%%%%%%%%%%%%%%%%%%%%%%
\acknowledgments

This research is supported by grants from the U.S. Department of Energy,
the U.S. National Science Foundation and the Smithsonian Institution, by
NSERC in Canada, by Science Foundation Ireland and by PPARC in the UK.
We acknowledge the excellent work of the technical support staff at the
FLWO and the collaborating institutions in the construction and
operation of the instrument. We acknowledge the efforts of the Swift
Team for providing the UVOT/XRT observations. We thank James Bedient of
the AAVSO for his V and I band data on W\,Com.

\end{document}